\documentclass[manuscript]{aastex}

\usepackage{emulateapj5}

\shorttitle{Scattered H$\alpha$ light from Galactic dust clouds}
\shortauthors{Mattila, Juvela \& Lehtinen}

\begin{document}


\title{Galactic dust clouds are shining in scattered H$\alpha$ light}


\author{K. Mattila \altaffilmark{}, M. Juvela  \altaffilmark{} and 
K Lehtinen \altaffilmark{}}
\affil{Observatory, University of Helsinki, T\"ahtitorninm\"aki, 
FI-00014 Helsinki, Finland}

\email{mattila@cc.helsinki.fi}






\begin{abstract}
Bright emission nebulae, or H{\small II} regions, around hot stars 
are readily seen in H$\alpha$ light. 
However, the all-pervasive faint H$\alpha$ emission has 
only recently been detected and  mapped over the whole sky.
Mostly the H$\alpha$ emission observed along a line of sight 
is produced by ionised gas {\it in situ}.
There are, however, cases where all or most of the
H$\alpha$ radiation is due to {\it scattering} by electrons or dust
 particles which
are illuminated by an H$\alpha$ emitting source off the line of sight.
Here we demonstrate that  diffuse, 
translucent and dark dust clouds at high galactic latitudes are in
many cases observed to have an excess of diffuse H$\alpha$ surface 
brightness, i.e. they are brighter than the surrounding sky. 
We show that the majority of this excess surface brightness can 
be understood as  light scattered off the interstellar dust grains.  
The source of incident photons is the general Galactic H$\alpha$
background radiation impinging on the dust clouds from all over the sky.
\end{abstract}


\section{Introduction}


Although in most instances the H$\alpha$ emission observed along a line of sight 
is produced by ionised gas {\it in situ} there are cases where part or all
of it is due to scattering.
In our Milky Way the  H$\alpha$ line radiation from the bright 
ionised regions in the mid-plane of the Galaxy is scattered off
the widely distributed dust and gives rise to a diffuse H$\alpha$ intensity
\citep{jura,wood}. 
Early attempts by  \citet{reynolds73} to detect the scattered H$\alpha$ light in high latitude
dust clouds have not been successful.

Localised interstellar dust clouds, especially small size globules, 
are frequently found as dark patches against the bright H$\alpha$ 
background in HII regions \citep{bok47}. However, the opposite phenomenon
of ``bright dark nebulae'' should also be observable because of the
scattered  H$\alpha$ light. This situation is expected to occur when
a dust cloud is projected towards a faint H$\alpha$ background
at high galactic latitudes. In this letter we show that the 
all-sky  H$\alpha$ map \citep{finkbeiner} reveals the presence of numerous
such clouds. 
Depending on the location of the ionised H$\alpha$ emitting gas along
the line of sight, i.e. the proportion of gas
in front of vs. behind  the dust cloud, the cloud may actually appear
brighter, darker or neutral relative to its surrounding  
 H$\alpha$ surface brightness.


\section{Observational data}

Using the all-sky  H$\alpha$ map \citep{finkbeiner} we have 
surveyed a sample of high and intermediate galactic latitude clouds 
listed in the catalogs of \cite{mbm} and \citet{reach}
and visible in IRAS 100 $\mu$m maps. 
We have found several clouds with an excess  H$\alpha$ surface 
brightness but also cases where the cloud appears in absorption or even cases 
with neither
``emission'' nor ``absorption'' even though the dust column density would predict
a clear effect. A list of  high galactic latitude clouds with an H$\alpha$ 
excess probably caused by scattering includes 
the following objects: the Lynds clouds L~79, L~134/169/183 complex, L~1407, L~1495, 
L~1532, L~1660, L~1780, the MBM objects \citep{mbm} MBM~16, MBM~46, MBM~137, MBM~145, 
and the diffuse infrared clouds \citep{reach} DIR~002+31, DIR~046+37,  DIR~340-43.
In Figure 1 we display the  H$\alpha$ and 100 $\mu$m images for a
selection of four high galactic latitude clouds together with the scatter
diagrams of  H$\alpha$ intensity vs.  100 $\mu$m intensity.

The detected excess emission is in the range of
1 to 3 rayleigh (R) which, when compared with the average all-sky
 H$\alpha$ surface brightness of $\sim$8 R, suggests that scattering
can be the source of this phenomenon.
The morphological appearance of the   H$\alpha$ intensity distribution  is very similar to the 
dust distribution as revealed by the IRAS 100 $\mu$m images (Fig. 1) and optical extinction maps. 
The scatter plots of $I_{H\alpha}$ vs.  $I(100 \mu m)$ (Fig. 1 right panel)
and  $I_{H\alpha}$ vs. $A_V$  (Fig. 2 left panel) show for the clouds L134/L183 and L1780,
which have low background  H$\alpha$ intensity,
a functional dependence characteristic of scattered light: first a linear increase
in the optically thin domain, $A_V \le 1$ mag, and a gradual saturation for the larger
optical thicknesses. Other clouds, appearing in ``absorption'' rather than in ``emission''
(Fig. 1 d; Fig. 2 c-e), can also be readily understood as combination of  H$\alpha$
scattering and extinction of  H$\alpha$ emission from beyond the cloud's distance.

The intensity of the  H$\alpha$ radiation along the line of sight 
towards an obscuring and scattering dust cloud at a distance $d$,
with the optical depth $\tau$ at 6563 \AA\,,
and with a geometrical depth through the cloud much smaller than the distance $d$,
can be expressed as

$I_{\rm H\alpha}^{\rm cl} = I_{\rm H\alpha}(r < d) + I_{\rm H\alpha}(r > d)\times e^{-\tau} + 
I_{\rm H\alpha}^{\rm sca}$ \hfill (1)

\noindent whereas in the surroundings of the cloud the intensity is

$I_{\rm H\alpha}^{\rm surr} = I_{\rm H\alpha}(r < d) + I_{\rm H\alpha}(r > d)$ \hfill (2)

\noindent and the difference is
 
$ \Delta I_{\rm H\alpha} = I_{\rm H\alpha}^{\rm cl} - I_{\rm H\alpha}^{\rm surr} =  I_{\rm H\alpha}^{\rm sca}- 
I_{\rm H\alpha}(r > d)\times (1 - e^{-\tau})  $ \hfill (3)

\noindent Thus, depending on the absolute values of the two terms on the right hand side the 
cloud's  H$\alpha$ surface brightness can be higher, lower, or equal to its surroundings. 

The concept of H$\alpha$
scattering is tested by comparing the observed H$\alpha$ surface brightness
with predictions of radiative transfer models.
We consider a spherical model cloud which is illuminated by the H$\alpha$ Interstellar
Radiation Field (ISRF).

\section{Radiative transfer modeling}

The simulations of scattered H$\alpha$ intensity were performed with Monte
Carlo methods  \citep{juvela03, Juvela2005a,mattila70b} using 
peel-off \citep{yusef94} methods to improve the sampling of the outcoming
scattered radiation.
The intensity of the background sky illuminating the
model cloud was obtained from the WHAM survey \citep{Haffner03} as
function of radial velocity. The missing southern sky was filled in by
mirroring the data with respect to the Galactic longitude.
The density distributions of the spherically symmetric model clouds were
calculated according to the Bonnor-Ebert solution \citep{bonnor,ebert} of hydrostatic
equilibrium with temperature 12 K and the nondimensional radius parameter $\xi$ = 7.0. 
The sizes and masses of the clouds were fixed by the visual extinction through
the cloud centre. Dust properties were based on Draine's (2003)
 `Milky Way Dust' model. 

During each run, photons of a single frequency were simulated and the
scattered intensity, including multiple scatterings, was registered as an
image of the cloud toward the observer. The full H$\alpha$ profile was covered at steps of
5\,km/s. The direction of the observer was determined by the actual galactic
coordinates of the cloud in question. This way, although the model clouds
themselves are spherically symmetric, the anisotropic radiation field can
produce asymmetrical surface brightness distributions. 
The resulting images of scattered H$\alpha$ intensity consist of 21$\times$21
independent pixels. The Monte Carlo noise, i.e., the rms-variation of pixel
values is about 1\%. 

The WHAM survey gives the distribution of the H$\alpha$ emission as seen from
the Earth. Although our sources are relatively nearby objects, the distribution
of the sky brightness, as seen from their location, can be  different. In particular,
part of the H$\alpha$ emission seen toward a cloud at a distance of $d$ may, 
in fact, reside between the cloud and us. Therefore, we have also examined scenarios, where a
given fraction of the H$\alpha$ emission 
seen in an area of the sky toward and around the cloud
is assumed to emanate in front of it. 
Thus, the H$\alpha$ intensity is 
divided into the components $I_{H\alpha}(r < d)$ and  $I_{H\alpha}(r > d)$ 
(see Eqs. 1 and 2). As seen from the cloud's location the emission
component  $I_{H\alpha}(r < d)$ is moved to the opposite part of the sky.
This modification is applied  to a region within 45\,degrees of the direction towards
the cloud where the fraction  $I_{H\alpha}(r < d)$ is subtracted from  $I_{H\alpha}^{surr}$.  
The modification is taken into account as an additional light
source in the opposite part of the sky when simulating the scattering of H$\alpha$ radiation. 
The surface brightness component $I_{H\alpha}(r < d)$ is not attenuated by 
extinction in the cloud.

The H$\alpha$ surface brightness behind the cloud, $I_{H\alpha}(r > d)$, 
 was originally set equal to the value observed in the direction of the cloud 
in the WHAM-survey. However, the resolution of that survey
is only one degree, and the surface brightness can have large variations
within each WHAM beam. Of course, the WHAM value also includes the object
under consideration
whereas in the modeling we need a value for the background without the
foreground source. Therefore, the background sky brightness was adjusted so
that its level corresponded to the background value observed in the higher
resolution H$\alpha$ map \citep{finkbeiner}. 


We examine in more detail the H$\alpha$ surface brightness of the clouds L~1780, 
L~1642, MBM~105, MBM~128/129, and MBM~144.
The clouds were modeled as Bonnor-Ebert spheres such that the range of extinction
corresponded to the observed values. 
There is only one free parameter that can be adjusted in order to find
the best correspondence between observed and modeled H$\alpha$ intensities,
namely the fraction $k_{\rm bg}$,  $0<k_{\rm bg}<1$, 
of the background H$\alpha$ intensity that resides behind the source, 
i.e.  $I_{H\alpha}(r > d) = k_{\rm bg}I_{H\alpha}({\rm surr})$. 
The remainder, $(1 - k_{\rm bg})I_{H\alpha}({\rm surr})$,  comes from the medium
between the cloud and the observer. 
Figure 2 shows the observed and modeled surface brightness as function
of visual extinction for four clouds.

  In the case of L~1780 the surface brightness
can be explained purely as scattered radiation. For the dust model 
\citep{Draine2003} used here the albedo is $a=0.67$  and the scattering 
function is  moderately forward directed with an asymmetry parameter
$g = 0.5$ at the wavelength of the H$\alpha$ line. 
In L~1642 neither an H$\alpha$ excess nor
absorption was observed. This can be explained by means of a suitable 
fraction of the H$\alpha$ emission coming from the medium behind the cloud.
In that case the scattered light from the cloud is compensated by
the attenuation of the background emission, $I_{H\alpha}(r > d)$.
 As a result, the surface
brightness variations remain very small. With $k_{\rm bg}=0.4$ the model
predicts a total variation of only 1\,R (see Fig. 2b).
In MBM~105, because of the
higher background level, the intensity decreases with $A_{\rm V}$.
The intensity towards the cloud centre is determined almost purely by the
scattering. In MBM~128/129 and MBM~144,  
with an even higher background intensity,
the importance of scattering decreases and the observations trace mostly
the extinguished background H$\alpha$ emission. 
When a value
$k_{\rm bg}=0.7$ is used, i.e., when some 30\% of the diffuse H$\alpha$
emission is assumed to originate in front of the clouds,
the models are in good agreement with the observations (see Figs. 2c and 2d).

Our model clouds are spherically symmetric. However,
 the impinging  H$\alpha$ radiation field is anisotropic. This means
that the intensity of the scattered light, as observed over the cloud's face, 
is no longer spherically symmetric.
The scattered light was registered as two-dimensional maps over the cloud's face. 
As an example, we
show in Figure 3 the calculated surface brightness maps for L~1780. 
The asymmetry is quite
noticeable at $A_{\rm V}=4^{\rm m}$. The lower frames show predictions for the
actual total surface brightness where the attenuated background component
has been added. Although this component is spherically symmetric it accentuates
the asymmetry of the scattered light by reducing the total range of surface
brightness values. The intensity maximum is clearly displaced towards the
south, i.e. towards the stronger radiation coming from the direction of the
Galactic plane.

\section{Discussion}

As shown in Figure 2 a the values of  
$a=0.67$ and $g=0.5$, valid for the dust model \citep{Draine2003} used in this study give a good 
agreement between the observed and calculated scattered light intensities for L~1780.
The difference can be easily explained, apart from a higher albedo value,
by a slightly higher intensity of the H$\alpha$ radiation field at the
location of the cloud, a value $k_{\rm bg}<1$, or inhomogeneity of the cloud
that would in the cloud centre cause a higher ratio between scattered
intensity and the line-of-sight optical depth.
For a given cloud direction $(l,b)$ a range of ($a,g$) combinations  
can reproduce the observed scattered light intensity. 

The asymmetry of the  H$\alpha$ intensity distribution is readily seen in
the observations of L~1780 and L~183: the H$\alpha$ intensity maximum 
is displaced toward the Galactic plane relative to the peak of the dust 
column density maximum (see Figs. 1a,b).
 This shift is $\sim$8 arc min in the case of L~1780.  

No other mechanisms except scattering are needed  to explain the faint 
H$\alpha$ excesses of 1-3 R as observed in high latitude translucent dust clouds.
The high latitude clouds are exposed to the ionising Lyman continuum photons
which escape from H{\small II} regions around O and early B type stars
near the Galactic plane. The   H$\alpha$ {\em in situ} emission expected
from  off-the-plane clouds has been estimated  using empirical
data for the distribution of hot stars and the intervening dust extinction
\citep{B98,B99a,BM02,putman}. For nearby clouds, $|z| < 500$ pc, the expected  H$\alpha$ 
surface brightness is $< 0.3$ R. On the cloud surface facing the
Galactic plane a thin layer of ionised hydrogen will form. Its thickness
is $\Delta s \approx 5$ 10$^{-2}$ pc/$n_ H$ which for an assumed hydrogen
density in the cloud's outer layers of $n_ H = 10 - 100$ cm$^{-3}$ corresponds 
to 5 10$^{-4}$ - 5 10$^{-3}$ pc. 
Morphologically, such a thin H$^+$ surface layer would be observed 
on the side facing the Galactic plane  as
uniformly distributed surface emission with no correlation with
the dust column density, and as a thin bright rim around the cloud's edge.
However, the angular thickness of such a bright rim would be of the order
1 - 10 arc seconds and could not be discerned by the present lower resolution
data  for L~1780, the L~134 complex (Figs. 1a,b), and other similar clouds.

The conclusion by \citet{delburgo} that the main mechanism
for producing the H$\alpha$ excess in 
L~1780 is the ionisation by an enhanced cosmic ray flux 
is not supported by our findings. 
However, some part of the emission could still come from such a process.

An important aspect
of the scattered H$\alpha$ line radiation as compared with the continuum radiation
is the velocity information it carries both on the source of the impinging
ISRF light as well as on the scattering dust particles 
\citep{B99b}.
Figure 4 shows
an example of the observed  H$\alpha$  scattered light line profile.
L~1780 shows an excess line profile centered at $V_{LSR} \approx$ 1 km/s
which differs from the mean velocity of the {\it in situ} H$\alpha$ emission in the
surroundings of the cloud, $V_{LSR} \approx$  -7 km/s. 
This is as expected since 
(1) the length of the velocity vector of the cloud itself is probably not much
larger than its radial velocity component, $V_{LSR}= 3.5$ km/s,
and (2) the impinging ISRF radiation
as integrated over the whole sky, has a mean velocity of  $\sim$ -1 km/s.
Model calculations as function of velocity of the impinging  H$\alpha$ radiation 
have been performed for L~1780. We have adopted 
three different scattering functions ranging from isotropic ($g = 0$) to a strongly
forward directed one ($g = 0.9$). While all the calculated 
scattered light  line profiles have closely the same mean velocity the line
width is smaller for the forward scattering as compared to the isotropic scattering
case. In the latter case the cloud ``sees'' with equal weighting the whole sky, 
with impinging H$\alpha$ velocities ranging from $V_{LSR} \approx$ -50 to +50 km/s.
The differences are, however, too small to allow a distinction between the
different values of $g$ (see Fig. 4).




\clearpage

\begin{figure*}
\includegraphics[width = 18cm]{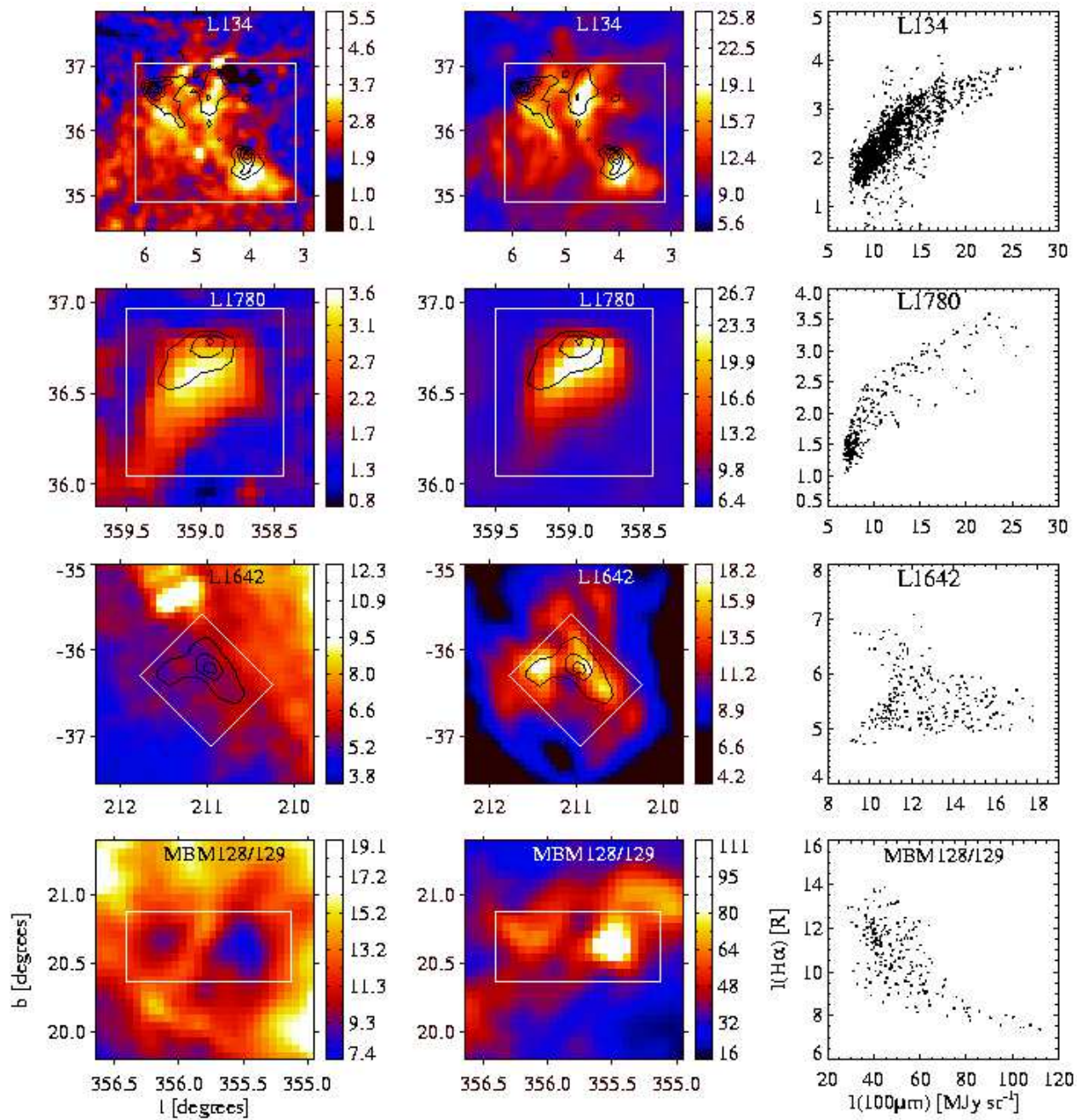}
\caption{{\bf Fig. 1} H$\alpha$ and 100 $\mu$m images  and  $I_{H\alpha}$ vs. $I(100 \mu$m) 
scatter plots towards some high-latitude molecular clouds. 
The leftmost column shows H\,$\alpha$ intensity in 
rayleighs (R) and the central column the 100\,$\mu$m surface brightness 
 in  MJy\,sr$^{-1}$. The intensity scales are indicated by the clour bars
on the right hand side of the images.  
The rightmost column shows pixel-to-pixel relation
between H\,$\alpha$ and 100\,$\mu$m intensities.
IRAS data have been extracted from the 
all-sky IRIS map  \citep{miville}. The pixel size in the images is
3.44$\times$3.44 arcminutes, which gives a pixel area that is equal to
the native pixel area of the HEALPix \citep{gorski} data used to make the 
maps.    The native resolution of the
H\,$\alpha$ data is 6\,arcminutes.  The resolution of the IRAS 100\,$\mu$m
data has been convolved into the same resolution. The axes are given in galactic 
coordinates. Notes on individual clouds:
{\bf (a) Lynds\,134 complex} consists of the
clouds L183, L169 and L134 (from upper left to lower right)
The contours show the extinction at $R$-band ($\lambda \sim$
650\,nm), derived via starcounts \citep{juvela2002}:
 1, 2, and 4\,mag.  This region is largely free of
foreground/background emission of H\,$\alpha$, thus the
morphological correspondence between scattered H\,$\alpha$
intensity and $R$-band extinction is good.
{\bf (b) Lynds\,1780} The cloud is seen in
H\,$\alpha$ excess emission against a dim H\,$\alpha$ background.  The
contours show the values of 200\,$\mu$m optical depth \citep{ridderstad}:
4\,10$^{-4}$, 8\,10$^{-4}$, and 12\,10$^{-4}$.  
{\bf (c) Lynds\,1642} This cloud is seen neither in excess emission nor in
absorption of H\,$\alpha$.  The contours show the values of
200\,$\mu$m optical depth \citep{lehtinen2004}:  
4\,10$^{-4}$, 8\,10$^{-4}$, and 12\,10$^{-4}$.  The
group of bright pixels at $l\sim 211.3^{\circ}$ , $b\sim-35.3^{\circ}$ in the 
H\,$\alpha$ map is an artifact caused by incomplete subtraction of bright stars.
{\bf (d) MBM128/129} This is an example of clouds seen in absorption against
a bright H\,$\alpha$ background. The two minima in the
H\,$\alpha$ map correspond to maxima in the 100\,$\mu$m map.}
\end{figure*}

\clearpage

\begin{figure}
\includegraphics[width = 10cm]{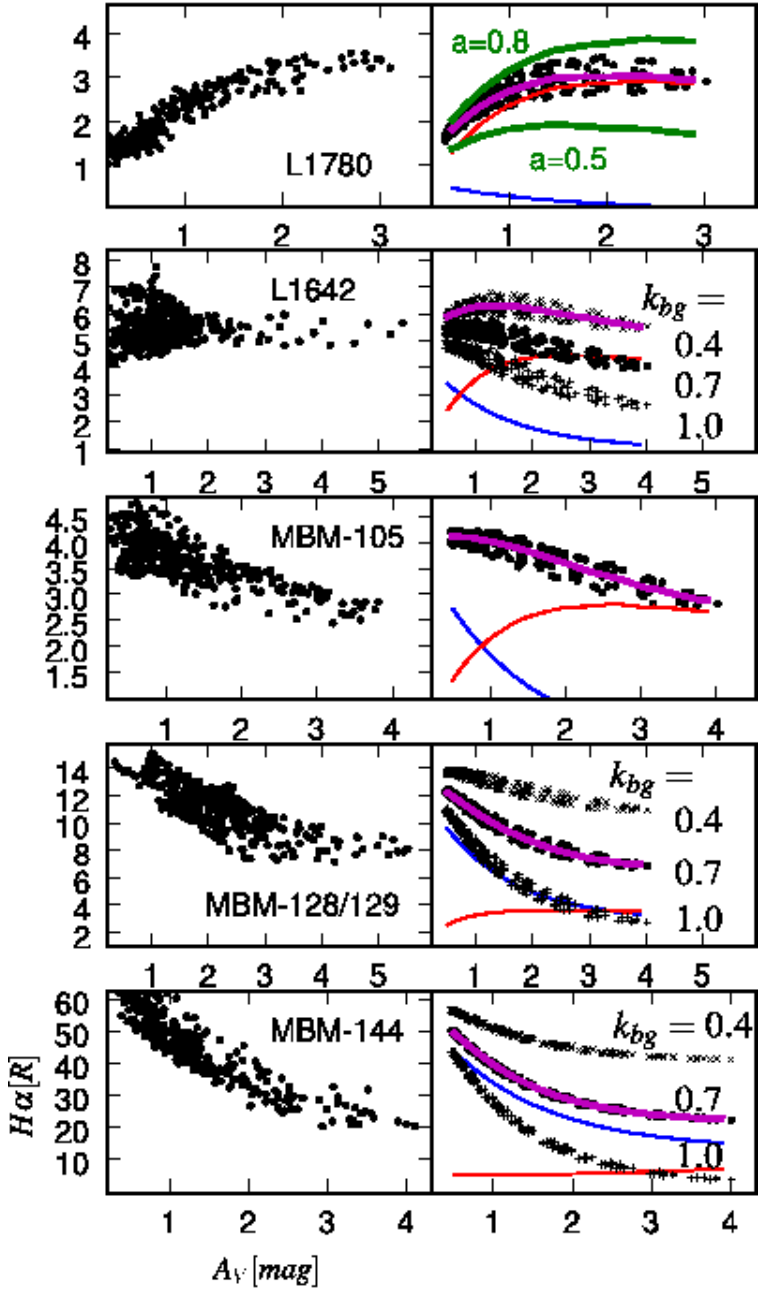}
\caption{{\bf Fig. 2}
Observed (left frames) and modeled (right frames)
 H$\alpha$ surface brightness as function of visual extinction. 
For L~1780 {\bf (a)} and MBM~105 {\bf (c)} the plotted model values correspond to
$k_{\rm bg}=1$. For L~1642{\bf (b)}, MBM~128/129{\bf (d)}, 
and MBM~144{\bf (e)} results are shown for three values of $k_{\rm bg}$. 
The magenta curves show the average  H\,$\alpha$ vs. $A_{\rm V}$ dependence of
the best fitting model. The red curves show the contribution of scattering,
and the blue curves the sum of the attenuated background intensity and, in
cases with $k_{\rm bg}<1$, the H-$\alpha$ emission that is assumed to reside
between the cloud and the observer. 
For L~1780 we also plot cases  $a=0.5$ or $a=0.8$.
The scatter of in the calculated intensities is caused mostly by the
anisotropic illumination which makes the surface brightness distribution
asymmetric (see Figure 3). A smaller part of the scatter is caused by the
Monte Carlo noise of the radiative transfer calculations.
The  $A_{\rm V}$ values were estimated using the reddening of the
background stars that were observed in the near-infrared 2MASS survey. The
colour excesses of the stars are converted into visual extinction using the
NICER method \cite{lombardi}. The resolution of the extinction map
was set equal to the resolution of the H$\alpha$ map, i.e., 6 arc minutes.}
\end{figure}

\clearpage

\begin{figure}
\includegraphics[width = 14cm]{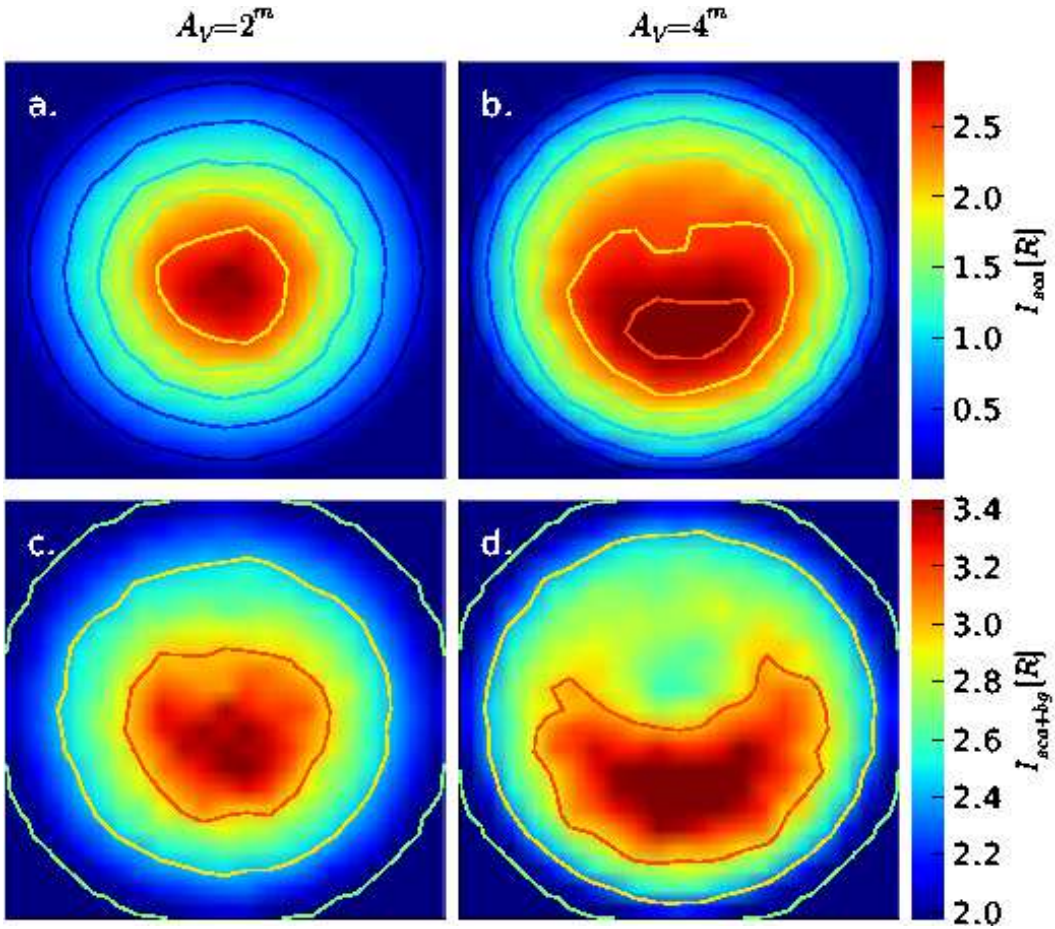}
\caption{{\bf Fig. 3}
Modeled H$\alpha$ surface brightness distribution in L~1780.
The maximum extinction through the cloud is assumed to be either $A_{\rm
V}=2^{\rm m}$ (frames {\bf a} and {\bf c}) or $A_{\rm V}=4^{\rm m}$ (frames
{\bf b} and {\bf d}). The upper frames show the distribution of the scattered
H$\alpha$ radiation, and the lower frames the total surface brightness where
the attenuated background intensity has been added (see Eq. 1). The contours are drawn at
intervals of 0.5~R. The asymmetry of the surface brightness is caused by the
anisotropy of the illuminating ISRF. The maps are oriented in Galactic
coordinates so that the radiation from the plane of the Galaxy comes from
below. The cloud size is specified only by its optical depth.}
\end{figure}

\clearpage

\begin{figure}
\includegraphics[width = 14cm]{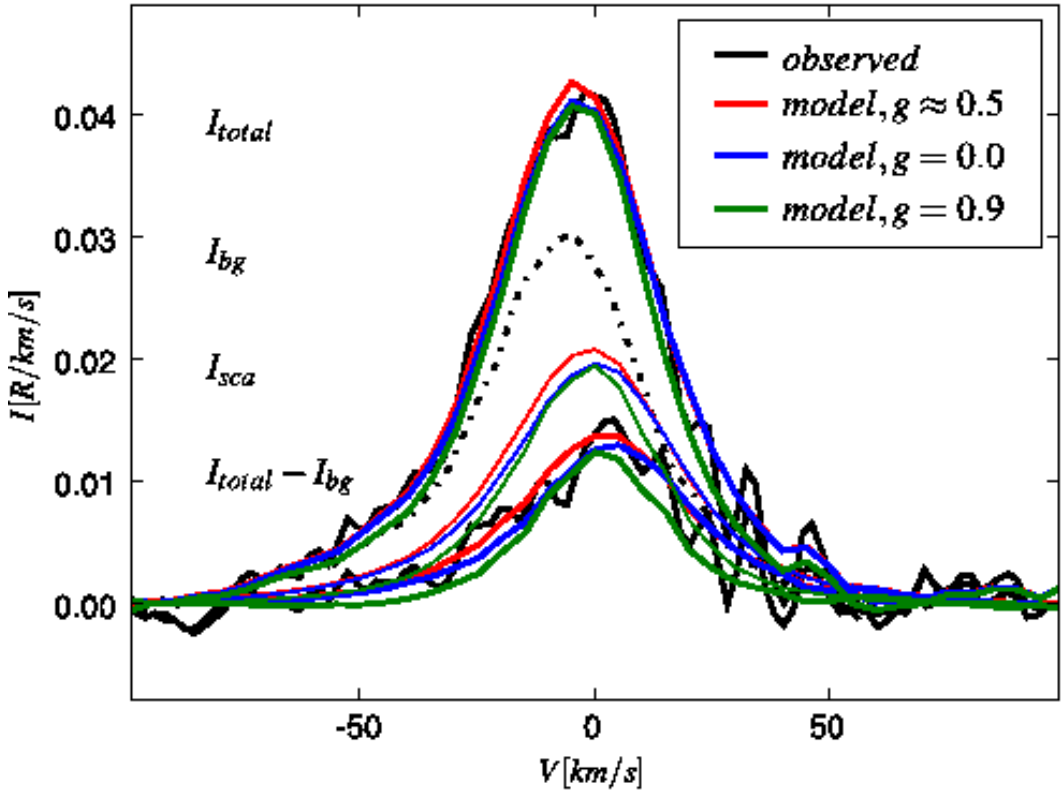}
\caption{{\bf Fig. 4}
Observed and modeled H$\alpha$ spectrum of L~1780.
 The figure shows the observed spectrum towards L1780
(uppermost black line), the average background around the cloud (black dash-dotted line), 
and their
difference (lower black line). The modeled average signal over the source ($I_ {\rm total}$) 
and the background-subtracted spectra ($I_{\rm total} - I_{\rm bg}$) 
are shown for Draine's (2003) dust model 
(red lines) and for modified dust 
models with isotropic scattering ($g=0$, blue lines) or strong forward scattering ($g=0.9$, 
green lines). The thin coloured lines indicate the corresponding spectra for scattered
light. The calculations are 
for a visual extinction of $A_{\rm V}=2^{\rm m}$ 
through the cloud centre.}
\end{figure}

\end{document}